# Damage Localisation in Fresh Cement Mortar Observed via In Situ (Timelapse) X-ray µCT imaging.

## Authors' Names and Affiliations

Petr Miarka[a,b*], Daniel Kytýř [c], Petr Koudelka[c], Vlastimil Bílek[d]

[a]Czech Academy of Sciences, Institute of Physics of Materials, Žižkova 22, 616 00 Brno, Czech Republic

[b]Brno University of Technology, Faculty of Civil Engineering, Veveří 331/95, 602 00 Brno, Czech Republic

[c]Czech Academy of Sciences, Institute of Theoretical and Applied Mechanics, Prosecká 76, Prague 9, Czech Republic

[d]VŠB — Technical University of Ostrava, Faculty of Civil Engineering, Ludvíka Podéště 1875/17, 708 00 Ostrava-Poruba, Czech Republic

*Corresponding Author:**

Petr Miarka (email: miarka@ipm.cz). Tel: +420 532 290 430

## Publishing information:

## Abstract

This paper presents the outcome of a study focused on the evolution of internal damage in fresh cement mortar over 25 hours of hardening. In situ timelapse X-ray computed micro-tomography (µXCT) imaging method was used to detect internal damage and capture its evolution in cement mortar hardening. During µXCT scans, the temperature released during the cement hydration was measured, which provided insight into the internal damage evolution with a link to a hydration temperature rise. The measured temperature during cement mortar hardening was compared with an analytical model, which showed a relatively good agreement with the experimental data. Using 20 CT scans acquired throughout the observed cement mortar hardening, it was possible to obtain a quantified characterisation of the porous space. Additionally, the use of timelapse µXCT imaging over 25 hours allowed for studying the crack growth inside the meso-structure including its volume and surface characterisation. The results provide valuable insights into cement mortar shrinkage and serve as a proof-of-concept methodology for future material characterisation.

## Highlights

- In situ timelapse X-ray µCT scanning of cement hydration.
- X-ray µCT analysis of crack growth in cement mortar.
- Measured and analytically calculated hydration heat.
- Porosity quantified using µXCT data.
- Crack propagation, volume, and surface in 3D from µXCT.

**Abbreviations**

| | | | |
|---|---|---|---|
| 2D/3D | two-/three-dimensional | CM | cementitious materials |
| AE | acoustic emission | MS | meso-structure |
| DIC | digital image correlation | OPC | ordinary Portland cement |
| μXCT | X-ray computed micro-tomography | ROI | region of interest |
| SCM | supplementary cementitious materials | | |

**Nomenclature**

| | | | |
|---|---|---|---|
| $D$ | specimen's diameter [mm] | $H_T$ | ultimate hydration heat [J/g] |
| $Cg$ | centre of gravity | $C_C$ | CM content [g/m$^3$] |
| $L$ | specimen's length [mm] | $t_e$ | equivalent time [hours] |
| $T$ | temperature [°C] | $t$ | time [hours] |
| $T_c$ | concrete temperature [°C] | $c_p$ | specific heat capacity [J·kg$^{-1}$·K$^{-1}$] |
| $T_r$ | room temperature [°C] | $E$ | activation energy [J/mol] |
| $\Delta T$ | temperature difference [°C] | $R$ | universal gas constant [8.314 J/mol·K$^{-1}$] |
| $Q_H$ | rate of heat hydration [J/m$^3$] | $\alpha(t_e)$ | degree of hydration at $t_e$ [-] |
| $Q_{X\text{-ray}}$ | heat generated by X-ray source [J/m$^3$] | $\alpha_u$ | ultimate degree of hydration [-] |
| $Q_{tot}$ | total heat [J/m$^3$] | $\beta$ | hydration shape parameters [-] |
| $h_1$ | height [mm] | $\rho$ | concrete density [kg·m$^3$] |
| $H$ | hydration heat [J/g] | $\tau$ | hydration time parameter [hours] |
| $H_{CEM}$ | total hydration heat of CM [J/g] | | |

# 1. Introduction

Regardless of the noticeable improvements in strength and the overall performance of modern concrete materials such as high-performance concrete (HPC) [1] and high strength concrete (HSC) [2], these are still sensitive to the formation of micro-cracks which can increase in size over the structure's lifespan. Modern concrete materials are being developed such that their production meets high environmental demands like $CO_2$ emissions [3–5] and a lower consumption of natural resources [6,7] (water, aggregates, additives) with the motivation of enhancing their mechanical performance and improving their durability, while reducing their overall cost.

Nonetheless, the ordinary Portland cement (OPC) is still being used either in its pure form, or, mostly, as a blended system [8] because of its reliable properties. OPC hydration is an exothermic process [9], and the total heat generated can have a significant influence on the in-place performance of concrete structures. Early hydration of OPC [10] and its modelling has been studied by various models [11].

HPC often has an improved permeability, apart from its other mechanical properties, which leads to its application in more hostile environments with a higher lifetime expectancy. Thus, cracks occurring in the protecting cover layer for steel reinforcement are more serious than in other applications, as it can have drastic consequences. The origin of these micro-cracks is in various load actions, e.g., temperature, static or cyclic loading, rheology of the concrete itself [12], and, especially, autogenous shrinkage [13,14]. Generally speaking, the mechanisms of concrete fracture are complex due to the highly heterogenous inner structure consisting of aggregates, cement paste, and air pores [15]. This complex behaviour is often studied by various fracture test setups.

In laboratory conditions, the concrete fracture process can be studied by numerous experimental methods, often under quasi-static loading, e.g., Moiré interferometry [17,18], acoustic emission (AE) [19], digital image correlation (DIC) [20–22], neutron tomography [23,24], and X-ray computed tomography [25,26]. The last-mentioned method – (μXCT) scanning – has especially attracted a growing interest [27] in material science and experimental mechanics research fields, as it allows for a non-destructive, high-resolution 3D analysis of material morphology, and internal damage localisation. The 3D model of the analysed samples is virtually reconstructed from 2D X-ray images/projections obtained at a different angle resulting in a spatial resolution, depending on the sample size, often less than 50 μm [28].

From the concrete technology viewpoint, μXCT allows for a precise measurement of internal flaws, e.g., pores [29], cracks [30,31], and inclusions [32]. More importantly, it allows for a comprehensive analysis of material heterogeneity [33]. Thus, the employment of μXCT scans during the loading of quasi-brittle materials is ideal for studying the damage evolution at the inner interface – at the aggregate/cement paste level. Moreover, μXCT could be used to study early age properties and pore structure [34,35], it could allow for a direct observation of $C_3S$ particles [36], and it could provide insights into the internal hydration behaviour [37].

Previous μXCT studies were focusing on the analysis of damage propagation in quasi-brittle materials under quasi-static [26,31,38,39] or dynamic [28] loading conditions. Such studies were capturing the μXCT scans while holding the specimens under a constant stress or constant deformation during the scanning process. This scanning method, unfortunately, cannot be extended to fresh cement mortars or other mortars due to a lack of solid phases, which can carry the load.

This experimental study presents a proof-of concept methodology for future material characterisation, which has placed focus on the damage detection analysis of early age behaviour of OPC mortar hardening and its natural process of heat release during cement mortar hardening. Cylindrical samples with cement mortar of water to cement ratio w/c of 0.3 were made, and the damage evolution in the meso-structure (MS) was observed by the in situ μXCT timelapse scanning setup. The specimens were attached with thermocouples to measure the temperature created by cement hydration to identify the hydration phases. Prior to μXCT scans, chemical composition and isothermal calorimetry of the OPC and mechanical strengths were obtained. The observed experimental results are linked to the damage present in the material's MS. The MS and the damage localisation were analysed by the μXCT technique at different times of OPC mortar hardening. Experimental results show the progressive damage in MS, which is governed mainly by the natural exothermic reaction of Portland cement hydration, and provides valuable insights into the processes governed by cement mortar shrinkage.

## 2. Materials and Experimental Details

In this section, we give a comprehensive overview of the studied material and the experimental methods used. First, we introduce the cement mortar mixture and the chemical composition of Portland cement. This is followed by the presentation of the geometry used. Afterwards, we present a μXCT test setup together with a temperature measurement and a method for porosity assessment. These methods provide a solid base for understanding the challenging problem of fresh OPC mortar hardening.

### 2.1 Mixture Composition

OPC was mixed with the natural sand 0/4 mm to create cement mortar. Portland cement CEM I 42.5 R was used as the binder with a constant dosage of polycarboxylate superplasticizer Glenium 300 (BASF, Germany). Sand and its sieve curve are presented in Figure 1.

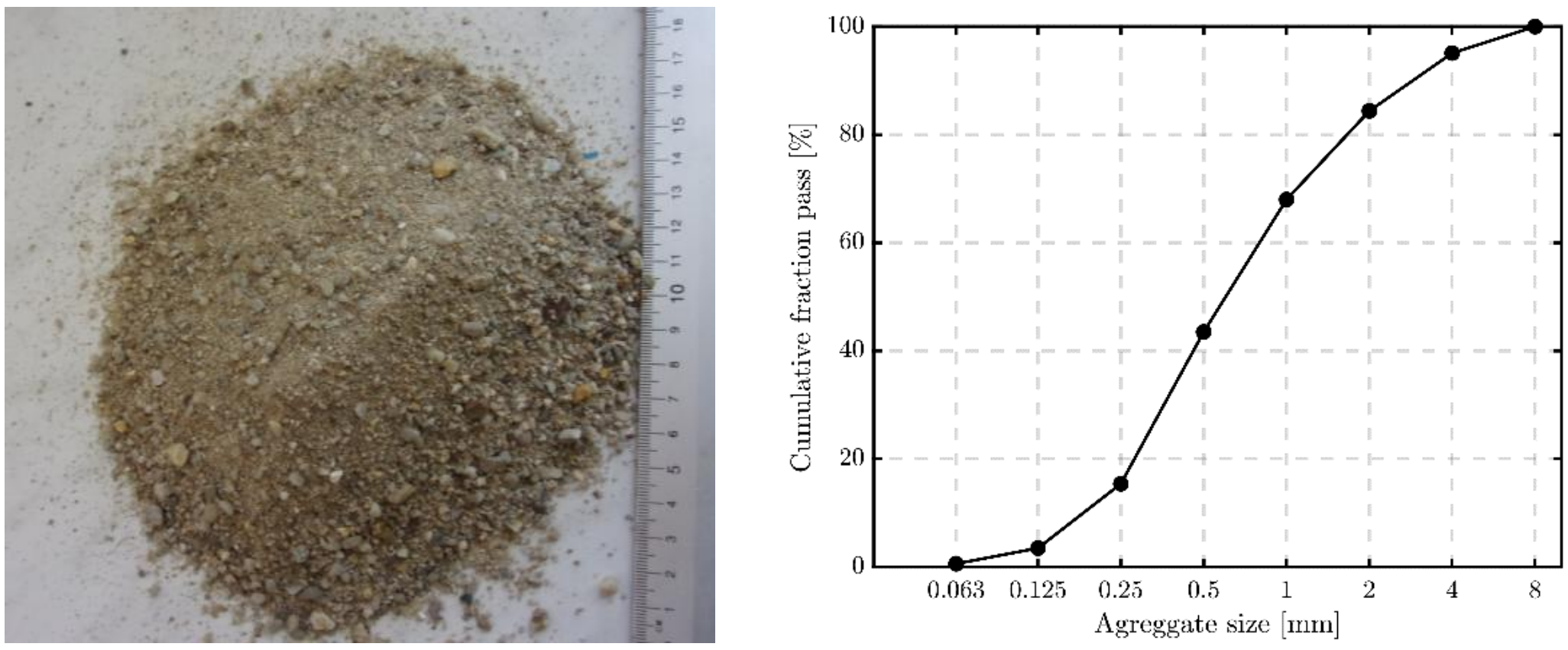

(a) (b)

**Figure 1:** Comparison of the used aggregates – (a) fine natural aggregate sand 0/4, and (b) sieve curve of the used sand.

The OPC mortar was mixed in a small volume batch of 1 l and poured directly into the plastic mould. The mixing, setting, and testing of mortar were performed in a room with a μXCT imaging setup (outside of laboratory conditions). The specimens were covered with a lid immediately after mixing and pouring to prevent from excessive water exchange with the environment.

This specific OPC mortar was selected, as it represents the mortar which would be present in HPC with w/c = 0.3. The coarse aggregates and water for pre-soaking of coarse aggregates were omitted. The absence of the coarse aggregates in the mixture serves to increase the spatial resolution of the μXCT imaging and to speed up post-processing of the obtained CT data. Thus, only three phase materials were analysed. The composition of the studied mixture per 1 l is shown in Table 1.

**Table 1:** Material composition of the cement mortar per 1 l.

| Volume | CEM I 42.5R [g] | Water [g] | Sand 0/4 [g] | Superplasticizer (Glenium 300) [g] |
|---|---|---|---|---|
| 1 l | 890 | 266 | 1 780 | 15 |

Table 1 shows the cement mortar composition per a small batch of 1 l, while the standard mixture composition would be per 1 m$^3$. This presentation is used to provide the most accurate data to the reader. But to provide a comparison with a standard mixture design, the cement content per 1 m$^3$ is 890 kg.

The mechanical properties presented in Table 2 were measured in accordance with European standards existing prior to the μXCT scanning procedure.

**Table 2:** Mechanical parameters (mean values with their standard deviations) of the cement mortar mixtures at various age days.

| Age [hours/days] | Volume density $\rho$ [kgm$^{-3}$] | Compressive strength $f_c$ [MPa] | Flexural strength $f_{c,t}$ [MPa] |
|---|---|---|---|
| 12 hours | 2351 ± 9 | 3.6 ± 0.1 | 0.6 |
| 24 hours | 2341 ± 12 | 35.3 ± 0.2 | 5.3 |
| 28 days | 2376 ± 12 | 85.4 ± 2.1 | 10.3 |

Table 2 shows compressive and flexural strengths at various ages in order to capture the strength development of the mortar. 12 hours strength was measured, because the mortar was hardened enough to allow for demoulding and could carry the load. Additionally, the 12 hours age is when the hydration of the cement is at its peak. Other strengths, namely 24 hours and 28 days, were measured to provide an overview of strength development in time and to bring a comparison with strengths seen in the literature.

Beside the mechanical properties, the chemical composition of the Portland cement CEM I 42.5R was measured, as the mineral content has a crucial role in the hydration heat development. The measured chemical composition of CEM I 42.5R is shown in Table 3.

**Table 3:** Chemical composition and clinker mineral fractions in percentage by weight (max. wt. %) of Portland cement CEM I 42.5R.

| $SiO_2$ | $Al_2O_3$ | $FeO_2$ | CaO | Free CaO | $SO_3$ | $K_2O$ | $C_3S$ | $C_2S$ | $C_3A$ | $C_4AF$ |
|---|---|---|---|---|---|---|---|---|---|---|
| 19.9 | 5.7 | 2.9 | 62.6 | 2.1 | 6.5 | 1.0 | 58 | 12 | 9 | 7 |

Table 3 shows the chemical composition of Portland cement CEM I 42.5R obtained by X-ray fluorescence (XRF) using XL3t – 980 (Thermo Fisher Scientific, USA). The chemical composition served as an input to the mineralogical model recently proposed by Shim et al. [40], which is a combination of a well-known Bogue's approach [41] with up-to-date findings. The clinker mineral fractions were calculated, and the results were compared with the standards.

Since Portland cement hydration is an exothermic reaction, the heat released during the reaction does not have a steady rate or a steadily changing rate. Thus, the hydration heat development can be separated into three hydration phases based on release rates [42,43]: (1) a short and immediate period of rapid heat decrease; (2) a longer period of steady temperature rise; and (3) temperature decline followed by a brief increase in heat emission before a stable steady decline to room or outside environment temperature. The first phase is the time when the dry cement first comes into contact with water and it corresponds to the initial hydration at the surface of the cement particles, mainly involving $C_3A$. This is followed by the so-called dormant period, at which the rate is zero, or nearly zero. The second phase typically has its peak at the age of 10 hours, which can change depending on the composition of the mixture. This is related to the connection between individual cement grains.

Correct hydration of OPC influences the future strength development. Thus, the hydration heat was measured in iso-thermal conditions of 20°C using calorimeter TAM AIR (TA Instruments, USA). Hydration heat was measured for two cement mortars (including cement and sand) with superplasticizer Glenium 300 and without it. The total weights of the mortars measured in the calorimeter were 10.83 g and 10.89 g, respectively. Hydration heat of the OPC mortars was being measured over a 7 day (168 hours) period, i.e., always when the heat of both samples reached a similar value. The measured hydration heat flow and cumulative heat is presented in Figure 2. Please note that the hydration heat is presented in double logarithmic coordinates as the time frame is long and the changes in the heat flow would not be visible in a standard coordinate system.

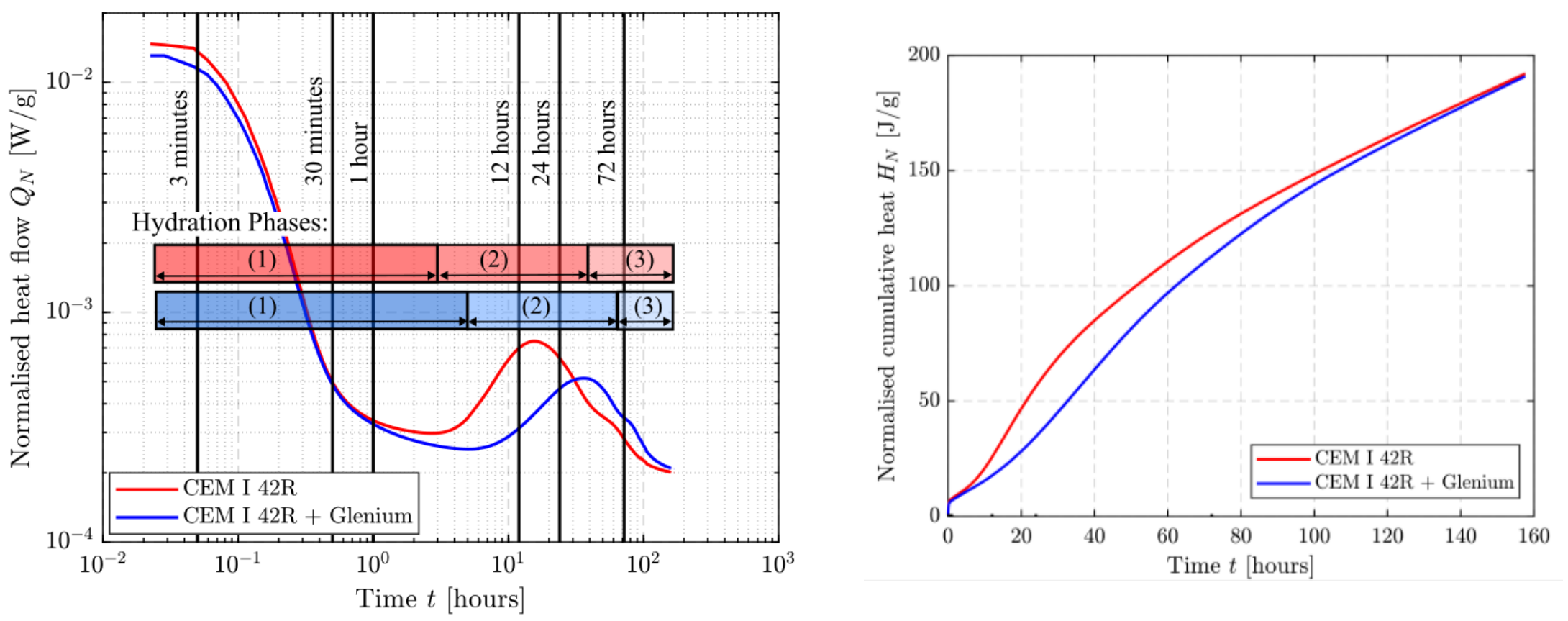

(a) (b)

**Figure 2:** (a) Iso-thermal hydration heat-flow, and (b) cumulative hydration heat of Portland cement CEM I 42.5R.

Figure 2(a) shows the hydration heat flow development of two OPC mortar samples with marked hydration phases. Both samples clearly have a different heat flow in the first two hydration phases. Such difference is related to the dosage of the plasticizer. Afterwards, the hydration heat flow shows a typical rapid decrease. The influence of superplasticizer is clearly visible, as the second peak in hydration heat is delayed to approx. 48 hours, while the plain sample evinces its peak at nearly 12:30 hours after mixing.

The difference in hydration heat of both samples starts at 0:30 hours when the sample without the plasticizer starts to produce heat (it slowly escapes the dormant period), whereas the sample with the plasticizer extends its dormant period until 6 hours from mixing. Thus, the difference in the dormant periods is approximately 3 hours.

From the cumulative heat release measured during the iso-thermal calorimetry as shown in Figure 2(b), the effect of superplasticizer can be observed. This slight difference in the measured heat curves starts again during the second stage of OPC hardening as the plain OPC sample starts this phase earlier and releases more heat than the sample with plasticizer. The measured cumulative heat curve serves as the input to the three-parameter hydration model [44], from which hydration temperature can be calculated.

## 2.2 Specimen's Geometry

In this experimental study, we selected geometry that allows for a reasonable spatial resolution of μCT images with a special attention paid to the container clamped on the rotary table of the CT system, i.e., low X-ray attenuation and geometry to allow for the acquisition of radiograms with a high signal to noise ratio for further volumetric data reconstruction and post-processing. Thus, we selected cylindrical specimens with a diameter *D* of approx. 29 mm with a length *L* of 140 mm. Moreover, the container's walls were selected to be X-ray transparent, i.e., made of a plastic material, in this case from polyethylene (PE). The dimensions of the cylinder are shown in Figure 3.

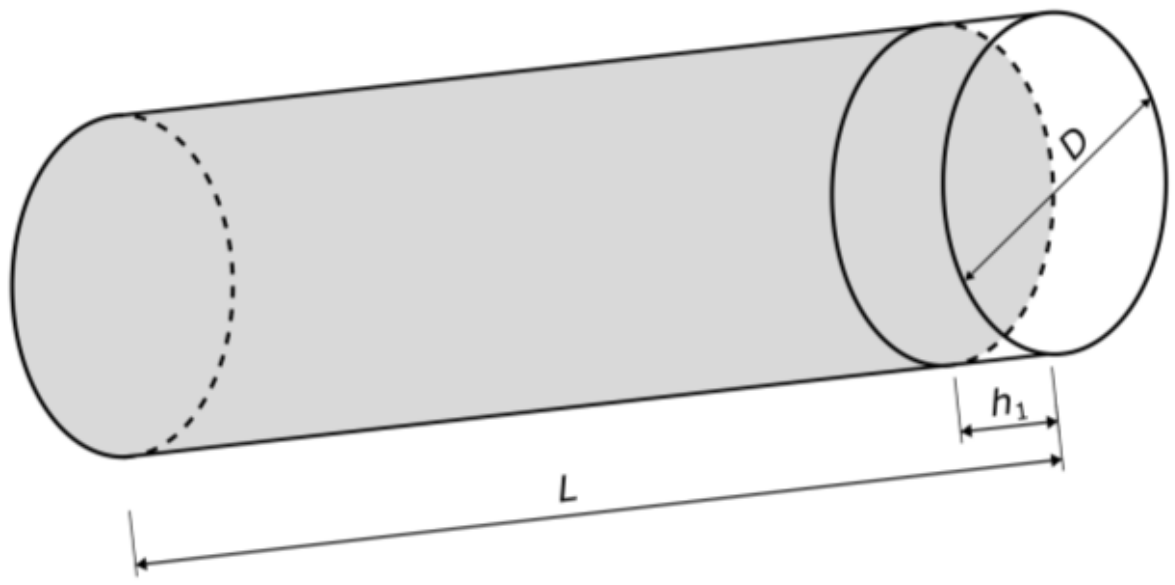

**Figure 3:** The geometry of the tested sample.

Figure 3 shows the sample geometry with the volume of the empty container with height $h_1$ on the right-hand side; in the experiment, this is the top surface of the sample. This volume was left empty to carefully insert the thermocouples used during the μXCT scanning process. The thermocouples were placed in the centre of the sample to measure the temperature produced by OPC mortar. Such positioning of the thermocouples allowed for axial rotation without redundant movement. Moreover, the thermocouples were placed outside the ROI. The thermocouple cable was placed in the middle of the lid as close to the centre of the sample as possible to provide accurate temperature data, while allowing for sample axial rotation without redundant thermocouple movement. In total, two thermocouples were placed in fresh OPC mortar to distinguish the sample's temperature development from the external heating by the X-ray source.

## 2.3 Cement Hydration Heat Model

The literature overview provides various analytical models for concrete hydration [11], which could be extended to the analysis of heat release influencing the stress of concrete structures structural members [45,46]. In this study, we used a hydration model of Schindler [44], which combines RILEM methodology [47] and experimental findings of Hansen [48].

The temperature development produced by a concrete specimen during the curing period under adiabatic conditions can be calculated as:

$$\frac{dT}{dt} = \frac{Q_H}{\rho \cdot c_p} = \frac{dH}{dt}\left(\frac{1}{\rho \cdot c_p}\right) \tag{1}$$

where *T* is the concrete temperature, $\rho$ is the concrete density, $c_p$ is concrete specific heat capacity, $Q_H$ is the rate of heat generation, and *H* is the heat of hydration of the concrete.

Temperature $T$ from Eq. (1) clearly depends on the heat generation rate $Q_H$, which could be determined from the following expression:

$$Q_H(t) = H_{CEM} \cdot C_c \cdot \left(\frac{\tau}{t_e}\right)^{\beta} \cdot \left(\frac{\beta}{t_e}\right) \cdot \alpha(t_e) \cdot \frac{E}{R}\left(\frac{1}{273 + T_R} - \frac{1}{273 + T_C}\right) \quad (2)$$

where $H_{CEM}$ is the total hydration heat of OPC, $C_C$ is the cementitious materials content, $\tau$ is the hydration time parameter, $\beta$ is the hydration shape parameter, $\alpha(t_e)$ is the degree of hydration at equivalent time $t_e$, $E$ is activation energy, $R$ is the universal gas constant (8.314 J/mol·K$^{-1}$), and $T_c$ and $T_r$ are the concrete temperature and reference (room) temperature, respectively. The rate of heat generation $Q_H$ is dependent on the degree of hydration $\alpha(t)$, which can be calculated from Eq. (3) as follows:

$$\alpha(t) = \frac{H(t)}{H_T} \quad (3)$$

where $H(t)$ is the cumulative heat of hydration released at time $t$ and $H_T$ is the ultimate hydration heat. Ultimate hydration heat $H_T$ can be calculated as $H_T = H_{CEM} \cdot C_C$. In this study, the Portland cement is used, and based on its chemical composition, the hydration heat of cement can be calculated. Thus, the hydration heat of cement $H_{CEM}$ will be used as follows:

$$H_{CEM} = 500p_{C_3S} + 260p_{C_2S} + 866p_{C_3A} + 420p_{C_4AF} + 624p_{SO_3} + 1186p_{FreeCaO} + 850p_{MgO} \quad (4)$$

where $H_{CEM}$ is the total heat of hydration of cement, and $p_n$ is the weight ratio of the $n$-th compound of the total cement content. Eq. (4) offers a possible adoption to the use of blended cements [44] with a content of supplementary cementitious materials (SCMs), e.g., slag, fly ash, metakaolin, etc.

The degree of hydration $\alpha(t_e)$, hydration time, and shape parameters $\tau$ and $\beta$ used in Eq. (2) are determined experimentally from adiabatic or semi-adiabatic calorimetry of hydration heat [47]. The experimental data can be fitted by an exponential curve as:

$$\alpha(t_e) = \alpha_u \cdot e^{\left(-\left[\frac{\tau}{t_e}\right]^{\beta}\right)} \quad (5)$$

where $\alpha(t_e)$ is the degree of hydration at equivalent age time $t_e$, and $\alpha_u$ is the ultimate degree of hydration. The ultimate degree of hydration $\alpha_u$ can be calculated based on water to cement ratio w/c as follows:

$$\alpha_u = \frac{1.031 \cdot w/c}{0.194 + w/c} \quad (6)$$

The ultimate degree of hydration $\alpha_u$ is unaffected by the concrete curing temperature, while it is strongly dependent on the water to cement ratio w/c. The role of water in the estimation of the ultimate degree of hydration $\alpha_u$ should include external water from the environment as well as the water present in aggregates [49].

## 2.4 X-ray µCT Test Setup

The radiographical imaging procedure was performed using a custom-built XCT laboratory system. For this experiment, µXCT is equipped with a 20W Microfocus X-ray source L10321 (Hamamatsu Photonics K.K., Japan), Dexela 1512 NDT (PerkinElmer, Waltham, MA, USA) flat panel X-ray detector with the active area of 1944 × 1536 px (75 µm pitch) using CMOS technology with Gadolinium oxysulfide scintillator, and rotary table ACD120-80 (Akribis Systems, Singapure) with +/-3 arc sec angular precision. The test setup for µCT is shown in Figure 4.

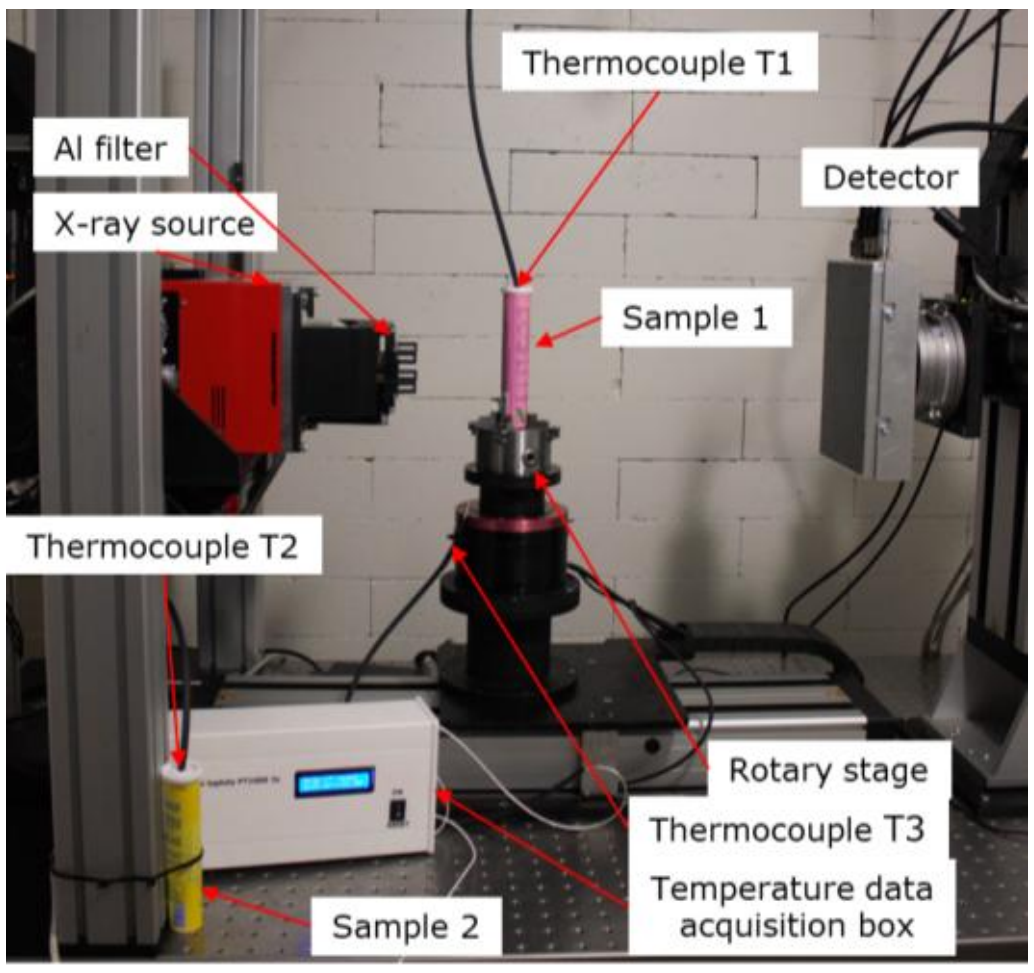

**Figure 4:** µXCT test setup with indication to cement mortar samples and thermocouples.

Figure 4 shows an over view of the experimental test setup used for µXCT data acquisition with marked scanned and reference samples. To acquire comprehensive information of heat generated during the cement hydration, Figure 4 shows high sensitivity thermocouples.

In total, three PT-1000 thermocouples (Kuongshun, China) with a range of -60°C ~ 200°C ± 0.05°C were used with the lengths of 1 m and 5 m. All thermocouples were connected using a MAX31865 amplifier to the Arduino Due board circuit used in the data acquisition. The circuit captures temperature data with a frequency of 1 Hz. The positioning of the thermocouples during the experiment was as follows: thermocouple T1) was placed in the scanned sample; thermocouple T2) was placed in the reference sample; and thermocouple T3) was mounted close to the scanned sample, but outside of the X-ray range to measure the room temperature.

For the scanning procedure, the acceleration voltage of 100 kV and a target current of 175 µA together with 4 mm Aluminium filter [50] was were used to obtain as high-quality an image of the radiograms as possible. Altogether, 20 time-lapse tomographic scans consisted of 1436 equiangular projections with 3×175 ms acquisition time. Including the positioning and read-out time, each tomographic scan took ~17 mins.

The reconstruction of individual 3D images was performed using a Feldkamp–Davis–Kress (FDK) filtered back projection reconstruction algorithm [51] implemented in VG Studio MAX 2023.2.1 (Volume Graphics, Germany) producing 3D data matrices with dimensions of 1943 × 1943 × 1536 voxels (voxel size of 20.6 µm). This provides a region of interest (ROI) with elliptical (circular) cross-section geometry of approximately 26.02 × 26.16 $mm^2$ by 36.14 mm; see Figure 5.

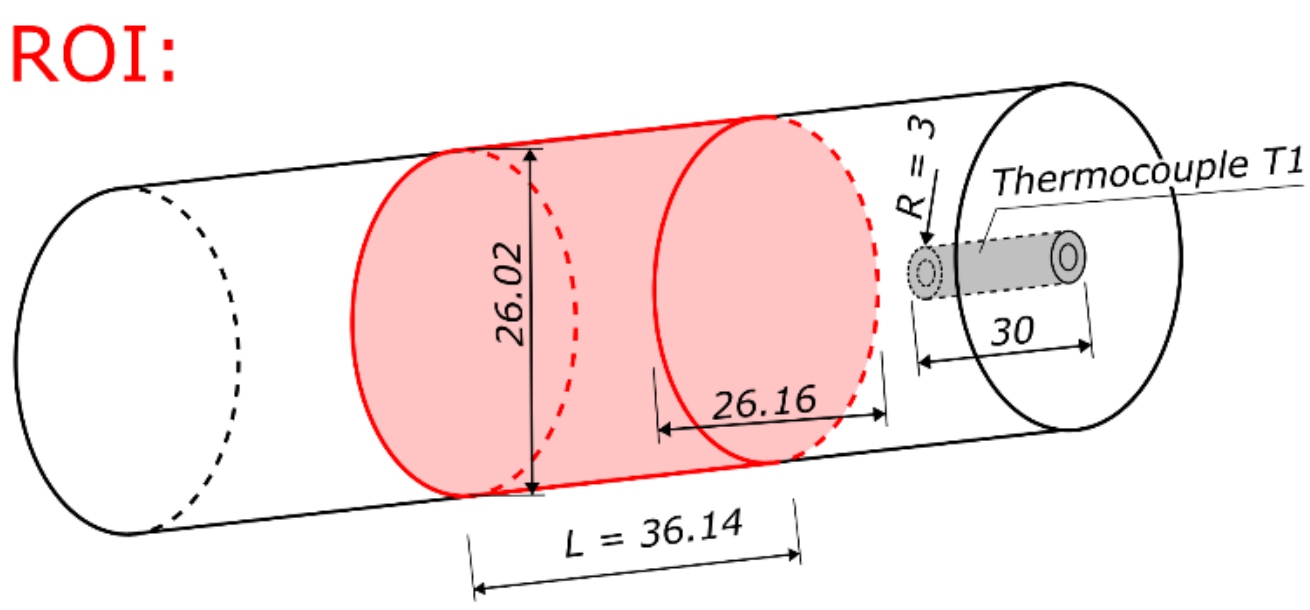

**Figure 5:** Schematic illustration of the analysed volume by µCT.

The ROI as presented in Figure 5 was selected to be in the middle of the cylinder outside of the possible reach of the inserted thermocouple. The size of ROI volume is 19 800 $mm^3$, which is later used in the porosity measurement.

## 2.5 X-ray μCT Data

Since the both concrete and cement mortar are multi-phase heterogenous materials, a threshold-based segmentation allows ~~to~~ for performing a phase separation to obtain separated 3D images of either the pore space, or the solid phases. The workflow of post-processing μXCT data is shown in Figure 6.

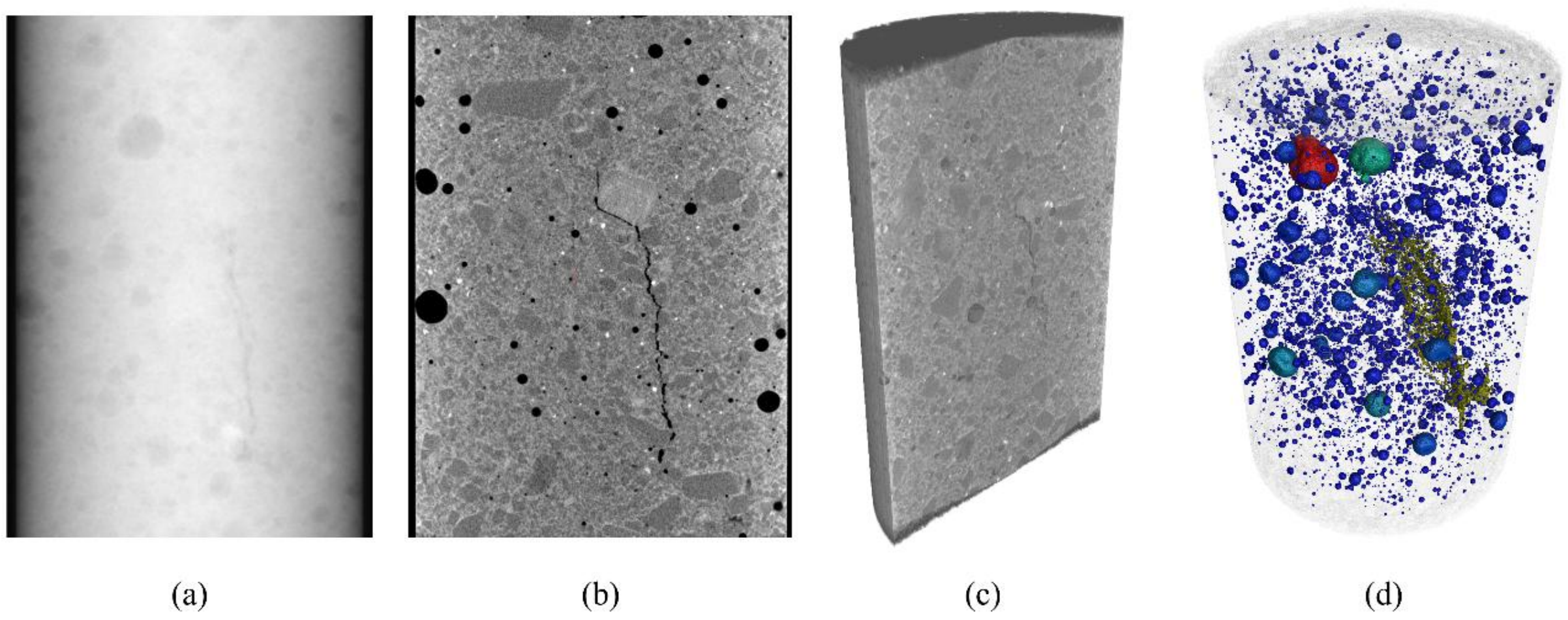

**Figure 6:** Illustration of the obtained X-ray data – (a) raw radiograph, (b) 2D view of CT reconstructed data, (c) 3D view of solid phases, and (d) separated pores.

Figure 6 depicts the workflow on the image analysis from the experiment to the identification of individual pores. During the experiment and the tomographic scanning of the sample, X-ray radiograms showing the specimen X-ray attenuation were acquired (Figure 6(a)). The set of equiangular projections was then reconstructed to a 3D image showing the specimen density, where Figure 6(b) shows a 2D view of a slice through such reconstructed volume including the pores and the crack. The density information contained in the reconstructed volume enables to study the material heterogeneity additionally to the pore space analysis using arbitrary visualisation means as depicted in Figure 6(c). Based on the procedure described in the following section, pore space and the crack can be quantified and visualised to, for example, the appearance in Figure 6(d), where the solid phases are suppressed in favour of individual pores and the crack coloured according to their calculated volume.

## 2.6 Porosity Measurement

To distinguish between pores and the emerging crack in the material, a porosity-space quantification was performed using tools implemented in the reconstruction software. In the first step, the surface of the specimen, i.e., the interface between the air and the solid, was identified on the basis of a gray-value threshold determined as an Otsu's threshold algorithm [52,53]. After the segmentation, we performed a pore identification procedure based on seed points for individual pore identification with the consideration of local gray value variation due to noise in the reconstructed 3D images.

The result of the analysis allows not only for studying local variations of porosity and porosity profiles along arbitrary directions in the reconstructed volume of the specimen, but also to for extracting every individual identified pore as a separate volume additionally to the calculated morphological characteristics. This approach was used to study and visualise the crack growth in the initial stages of the experiment. Here, the pore and crack localisation parameters were set on the first CT data post-processing and kept the same throughout the whole experiment. This, coupled with stable long term emission characteristics of the X-ray source and the response of the X-ray detector, ensured the highest possible reliability of the porosity quantification.

## 3. Experimental Results and Discussion

In this section, a comprehensive overview of the obtained experimental results is given. We begin with the presentation of measured the temperature development during mortar hardening from both samples. Afterwards, we present an internal damage analysis using µXCT scans in detail. The damage detected in the material's meso-structure is linked to the measured temperature profile.

### 3.1 Hydration Temperature Development

In order to link meso-structural changes due to the internal processes related to OPC mortar hardening, the temperature of the mixture was measured during the experiment over a time period of 25 hours. The captured temperature data provide comprehensive information on the hydration process and allowed us to distinguish from the external heat sources. The temperature captured during the µXCT scanning experiment is shown in Figure 7.

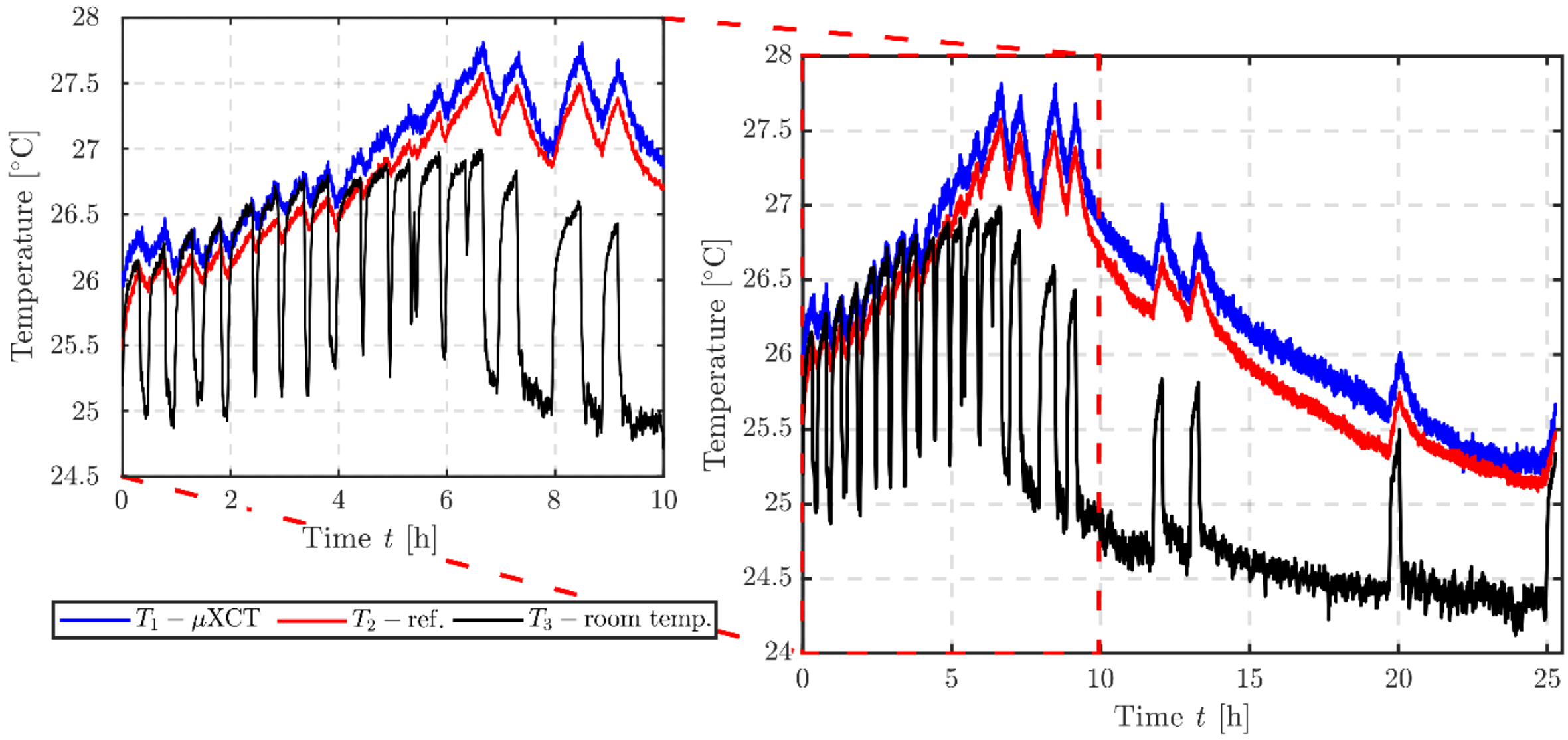

**Figure 7:** Temperature development over 25 hours of cement hydration with a focus on the first 10 hours.

Figure 7 shows raw temperature data acquired from the T1-T3 thermocouples during the first 25 hours of OPC mortar hardening during the µXCT scanning experiment. The peaks in the temperature data are related to the start of the µXCT data acquisition, as the X-ray source produces heat, which results in an increase in the room temperature due to the closure of the X-ray shielding door and its latter opening, respectively. The temperature data for both samples, reference and scanned, have a similar trend in the first 8 hours. After this period, the measured temperatures diverts from each other with a slight temperature increase in the scanned sample, keeping this trend until the end of the experiment.

The well-known cement hydration phases are more visible when the room temperature data $T_R$ (data from $T_3$) is subtracted from the measured sample's temperature data $T_C$ of concrete samples (data from $T_1$ and $T_2$) as $\Delta T = T_C - T_R$. The temperature difference $\Delta T$ then provides a better insight into internal temperature development of both samples and allows us to distinguish it from the external heat generated by the X-ray source. The temperature data were smoothed using a moving average filter with a moving window of 2.5 hours as implemented in the MATLAB 2023a (MathWorks, USA) software. The mean temperature difference $\Delta T$ data together with the original data from both samples are presented in Figure 8.

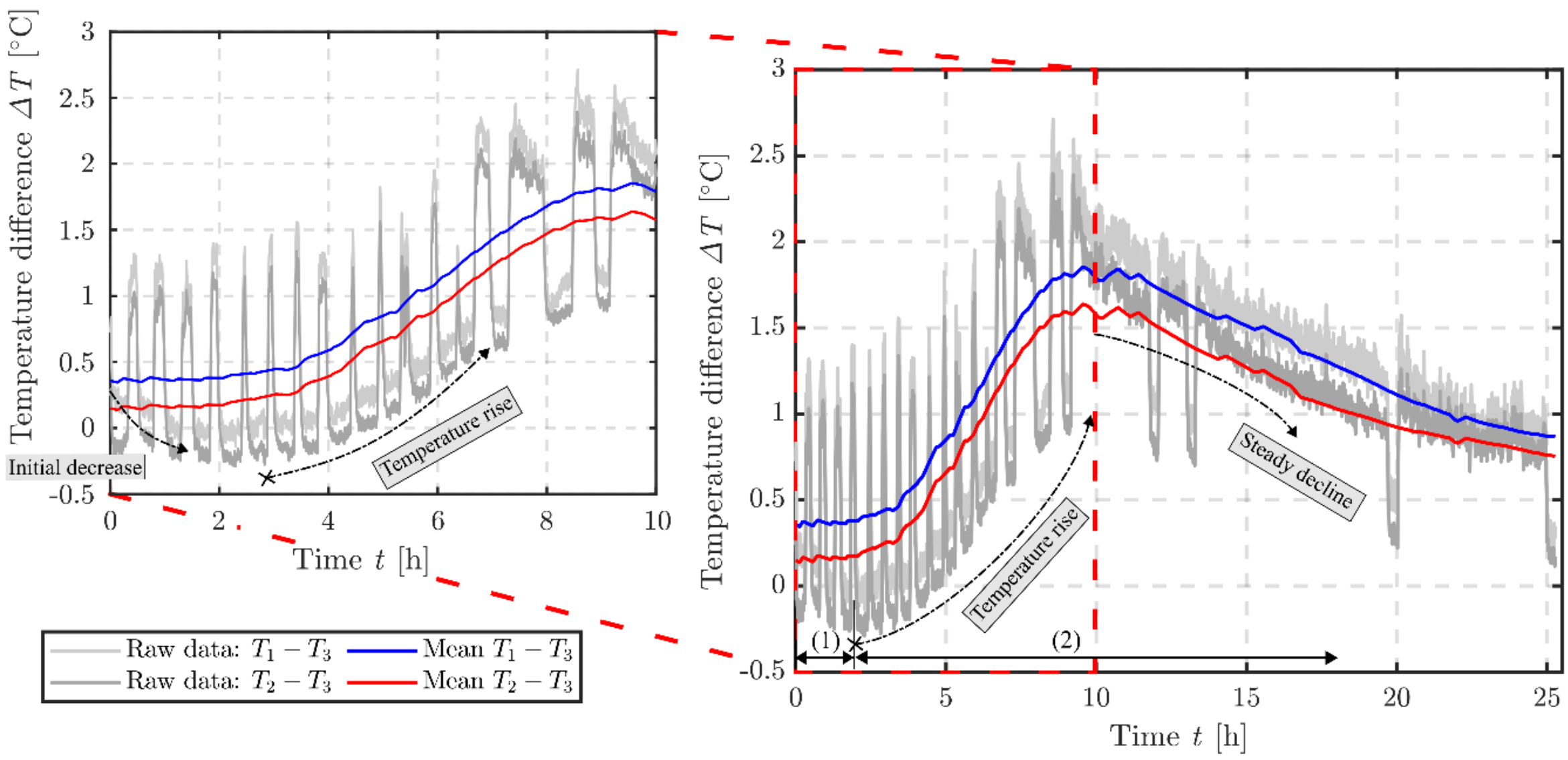

**Figure 8:** Mean temperature difference $\Delta T$ data with the focus on the first 10 hours with indicated hydration phases.

From Figure 8, a clear difference in the measured temperature can be observed between the thermocouples T1 and T2. This difference is approximately 0.3 °C and is again related to the heat produced from the X-ray source since the reference sample was place outside of the X-ray source range. Please note that the temperature profiles as presented in Figure 8 cannot be compared with the iso-thermal hydration temperature data shown in Figure 2 as the thermal conditions were different during the experiments (this includes the X-ray heat source and the consequent room temperature increase).

A careful observation of the raw temperature data as presented in Figure 8 shows a short decrease in the temperature in the period from 0:00 hours to 0:40 hours after mixing. This could be related to the first phase of cement hydration mentioned above. Unfortunately, the initial rapid decrease was not fully captured due to the sample preparation in another laboratory room (the transfer of the mixture from the mixer to the moulds and clamping of the sample in the rotary table). This took us approx. 10 minutes.

To verify the measured temperature with the analytical prediction as presented in Eq. (1), one needs to acquire chemical composition of the Portland cement. In this study, we adapted data available in the literature [44], which is similar to Portland cement CEM I 42.5R type. The adopted chemical composition was measured according to the ASTM standard, and is shown in comparison with the current OPC in Table 4.

**Table 4:** Clinker mineral fractions of current OPC compared with ASTM Type I cement adapted from [44].

| Type | $SiO_2$ | $Al_2O_3$ | $FeO_2$ | CaO | Free CaO | MgO | $SO_3$ | $Na_2O + 0.658K_2O$ | $C_3S$ | $C_2S$ | $C_3A$ | $C_4AF$ |
|---|---|---|---|---|---|---|---|---|---|---|---|---|
| 1 | 19.9 | 5.7 | 2.9 | 63.6 | 2.9 | 1.3 | 3.5 | 0.69 | 57 | 14 | 10 | 8 |
| 12 | 20.9 | 5 | 1.8 | 65.4 | 1 | 1.4 | 2.9 | 0.52 | 63 | 12 | 10 | 6 |
| 13 | 20.1 | 5.3 | 3.2 | 65.5 | 0.8 | 0.6 | 3.3 | 0.67 | 64 | 9 | 8 | 10 |
| Current | 19.0 | 5.7 | 2.9 | 62.6 | 1.0 | - | 6.5 | 1.0 (only $K_2O$) | 58 | 12 | 9 | 7 |

The clinker mineral fractions of OPC as presented in Table 4 show a similar content of $C_3S$, $C_2S$, and $C_3A$ when compared to the data provided by Schindler [44] with the difference limited to 6 %. Other parameters used in Eq. (2) are presented in Table 5Table 5 (please note that these parameters are based on a semi-adiabatic process).

**Table 5:** Cement hydration model parameters adapted from Schindler [44].

| No. | Cement type | *w/c* | *E* [J/mol] | Hydration parameters | | | $H_{CEM}$ [J/g] |
|---|---|---|---|---|---|---|---|
| | | | | $\beta$ | $\tau$ | $\alpha_u$ | |
| 1 | Type I: Source A | 0.37 | 45 991 | 0.905 | 13.69 | 0.689 | 477 |
| 12 | Type I: Source B | 0.50 | 41 977 | 0.719 | 16.88 | 0.887 | 513 |
| 13 | Type I: Source C | 0.50 | 46 269 | 0.727 | 16.32 | 0.882 | 492 |

Table 5 provides fitting parameters of three cement types based semi-adiabatic calorimetry adopted from Schindler [44]. Table 5 provides w/c ratios of all mixtures which are higher than the one in this study. Thus, we have decided to fit Eq. (5), which will provide more accurate parameters for temperature predictions within this experiment.

By substituting clinker fractions from Table 4 to Eq. (4), we can obtain the value of hydration heat $H_{CEM}$, which is 494.06 J/g. Moreover, if Eq. (6) is used with the current w/c = 0.3, we can obtain the ultimate degree of hydration $\alpha_u$ of 0.6261. These fitting parameters $H_{CEM}$ and $\alpha_u$ are within the range of reference cements. The fitting parameters of hydration temperature development as presented in Eq. (5), were obtained by the least square method and are presented in the following Table 6:

**Table 6:** Obtained fitting parameters of a three-parameter model.

| Cement type | *w/c* | *E* [J/mol] | Hydration parameters | | | $H_{CEM}$ [J/g] |
|---|---|---|---|---|---|---|
| | | | $\beta$ | $\tau$ | $\alpha_u$ | |
| OPC mortar | 0.3 | 40 150 | 0.884 | 30.97 | 0.6261 | 494.06 |
| OPC mortar + Glenium | 0.3 | 40 150 | 1.446 | 43.11 | 0.6261 | 494.06 |

Hydration parameters $\beta$ and $\tau$ as presented in Table 6 are within the range of the parameters found in the literature [54–56] for OPC mortar. The fitted parameters for OPC mortar with superplasticizer Glenium 300 are outside the parameters found in literature as the most of the studies used plain OPC mortars. To better illustrate the fitted parameters and their accuracy, a side-by-side comparison of the experimental and the fitted cumulative heat curve is presented in Figure 9.

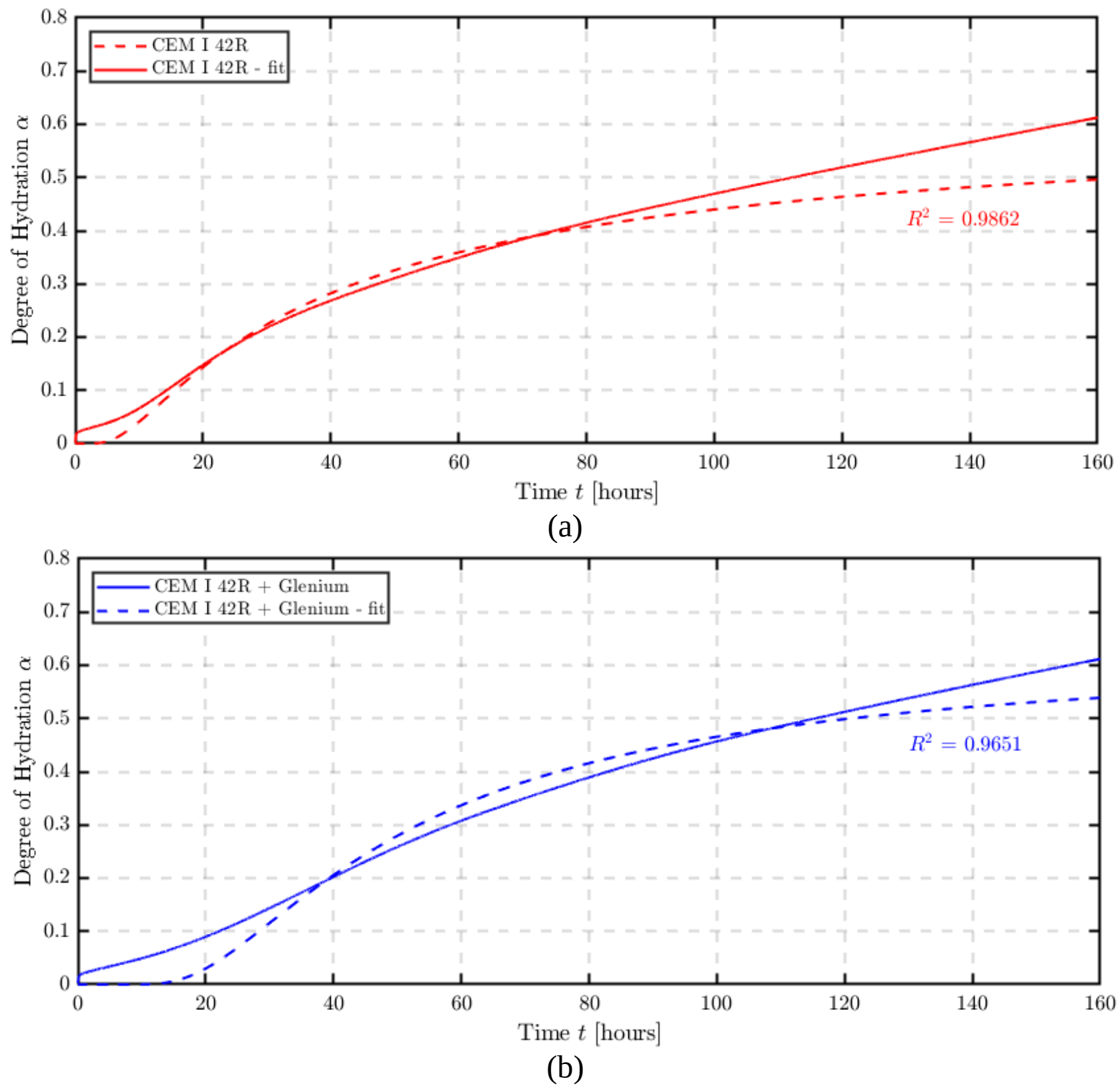

**Figure 9:** Comparison of fitted cumulative heat curve of analysed OPC mortar – (a) plain OPC mortar, and (b) OPC mortar with plasticizer Glenium 300.

Both fitted cumulative heat curves as presented in Figure 9 show a high accuracy of fit with the coefficient of determination $R^2$ of 0.9862 and 0.9651 for plain OPC mortar and mortar with superplasticizer, respectively. After a careful investigation of Figure 9(b) one can notice a delay in the onset of the degree of hydration for the fitted curve of the OPC mortar with Glenium 300. This observed delay of approximately 20 h in the increase in the degree of hydration again agrees with the general expectation that Glenium 300 acts as a retardant in OPC mortar. However, this delay does not agree with the measured experimental temperature data, as presented in Figure 9.

In this study, we used the above-presented three cement types (Table 4), as they show a similar mineralogical content in comparison to OPC used in this study as well as the fitted parameters from Table 6.

In the analytical prediction of hydration temperature as presented in Eq. (2), concrete temperature $T_C$ was set to 28 °C (the measured maximum), and room temperature $T_R$ was equal to 23°C. Cement content $C_C$ was $890 \cdot 10^3$ g/m$^3$, which is equivalent to the amount used in the our mortar. The concrete density $\rho$ and heat conductivity $c_p$ are 2400 kg·m$^3$ and 837 J/kg·K$^{-1}$, respectively.

The target power of the X-ray source of approx. 17.5 W was added to the total heat produced by the sample. Thus, the total heat considered in the analytical prediction corresponds to $Q_{tot} = Q_H + Q_{X\text{-ray}}$, and it was used in Eq. (1) to predict the temperature development of the sample. The comparison of the measured and calculated temperatures of the samples is presented in Figure 10.

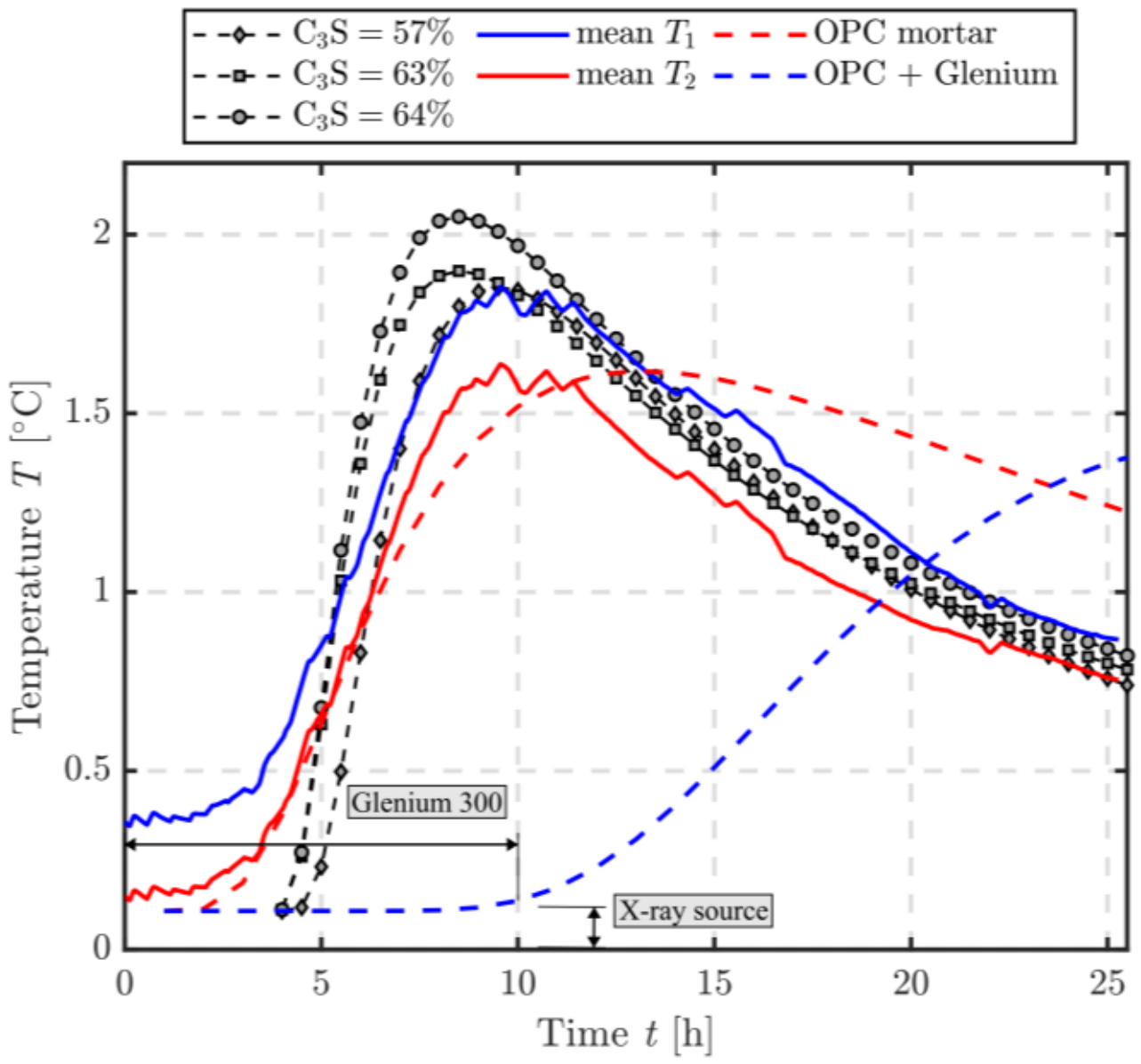

**Figure 10:** Comparison of measured and analytically predicted cement hydration temperature of the sample.

An interesting observation in Figure 10 can be made between the experimental and analytical hydration temperatures, as the analytical prediction is capable of capturing the experimentally measured temperature with a relatively good agreement. Additionally, Figure 10 shows the dependency of hydration heat on the $C_3S$ content. The predicted temperature for 57 % of $C_3S$ is capable to of predicting the temperature peak during the hydration heat most accurately, while other curves for 63 % and 64 % of $C_3S$ only show a similar trend with a reasonable error in the peak temperature. Moreover, the predicted temperature is capable of including the above-discussed external heating of the X-ray source, in which case it is approximately 0.3 °C. The delay in time in the hydration temperature rise in phase (2) is related to the use of the plasticizer Glenium 300, which delays the onset of hydration in the second phase by approx. 3.5 hours as observed in iso-thermal calorimetry.

Furthermore, the analytical temperature curve of plain OPC mortar is able to predict the temperature rise more accurately, which is more critical in this study than the temperature after the hydration peak, when compared with the curves based on the data from the Table 5 (Schindler et al. [44]). Additionally, the temperature curve of OPC mortar with superplasticizer Glenium 300 does not correspond to previously observed behaviour in isothermal calorimetry. The delay in temperature rise observed in the analytical prediction is nearly 10 hours, as the primary focus of this study was on the early age of cement hydration.

We attribute these differences between isothermal calorimetry and experimentally measured temperature profiles to the heating of external sources presented in the room with the CT setup. Nonetheless, this relatively good agreement between the analytical and experimental hydration temperature development as presented in Figure 10 serves as a clear link to the meso-structural analysis and further damage localisation, and it verifies the behaviour of OPC.

## 3.2 Porosity Measurement

Using the implemented algorithms that use gray intensity contrast-based algorithms and knowing the size of ROI, a porosity of the sample can be calculated. The combined instrumentation used in the experiments and the software tools are able to detect pores with a size bigger than the spatial resolution of 20.5 μm/Voxel. Since particle size distribution (PSD) of concrete can vary from mm to nm [57], we consider porosity as air voids, which are formed as air is entrapped during mixing. Such pores can have a size ranging from mm to μm, which could be easily detected by the achieved spatial resolution. The pores with a size smaller than the spatial resolution are considered to be noise.

The typical results of porosity analysis are the total pore volume, from which the porosity is calculated. In this study, the porosity measurement was done for every acquired μXCT scan, i.e., 20 porosity calculations. The calculated sample's porosity over time is presented in Figure 11.

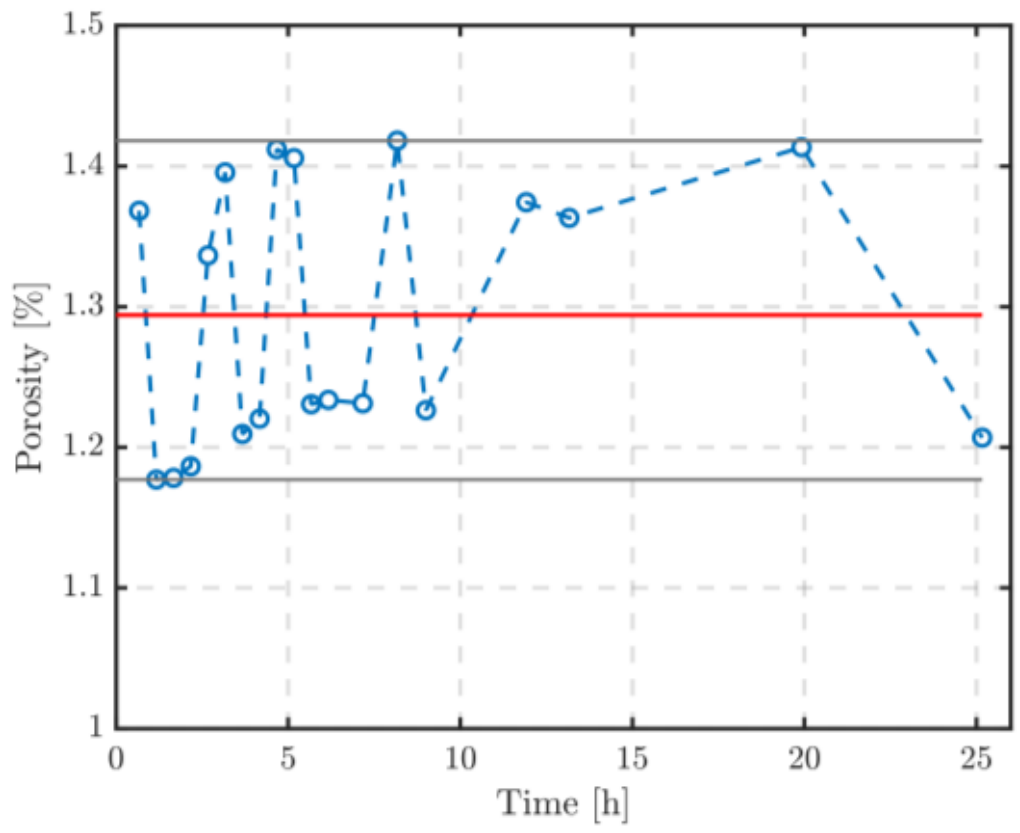

**Figure 11:** Porosity measurement during the μXCT.

The initial investigation of the porosity from CT data analysis in Figure 11 indicates a high scatter changing with nearly every CT scan. However, after performing simple statistics, the mean porosity is 1.29 % ± 0.12 %, which is in compliance with the results presented in Figure 11. It should be noted that the final value of mean porosity was calculated without the consideration of the volume of the macroscopic crack, i.e., the presence of the crack does not have any influence on the calculated average porosity (see the next subsection).

The observed scatter in porosity at an early age (up to 10 hours) could be linked to the hydration/hardening process as the wet cement mortar is drying in the container and is consuming the water from the mixture. Consequently, the OPC hydration is changing the material's density, which leads to variations in contrast-to-noise ratio in the reconstructed 3D images. As a result, the uncertainty in the determination of the material-air boundary increases and influences the thresholding procedure for the determination of the material iso-surface prior to the pore-space characterisation. This results in changes of the material density in the threshold values difference. Nonetheless, the measured porosity throughout the μXCT experiment can be considered an acceptable result as the error is < 0.2%, which, for heterogenous materials, is very low.

The identified pores appear in the samples during pouring into the mould and handling the sample, which agrees with the general expectation as the pores do not arise due to an inner chemical reaction during cement hydration. The obtained values of porosity correspond to the values found in the literature [58,59] ranging from 1% to 4% depending on the type of concrete. Moreover, the amount of entrapped air in HPC with w/c of 0.3 and with superplasticizer is usually from 1-1.5% [43]. The size of the observed air voids in the hardened state would not have a major influence on the material performance, e.g., compressive strength and Young's modulus, as the fluctuation of porosity is negligible.

### 3.3 μXCT – Damage Analysis

Using the same approach as for the porosity measurement, the damage (cracks) present in the sample could be identified throughout the experiment. In this case, a major crack was detected at 00:40 hours from mixing the samples, which increased in size until the experiment time of 06:10 h. After that, the change in crack size was insignificant for the analysis. The μXCT damage analysis provides valuable information on the crack surface and crack volume development during OPC mortar hardening. The obtained crack surface is presented in Figure 12(a), while the crack volume is shown in Figure 12(b).

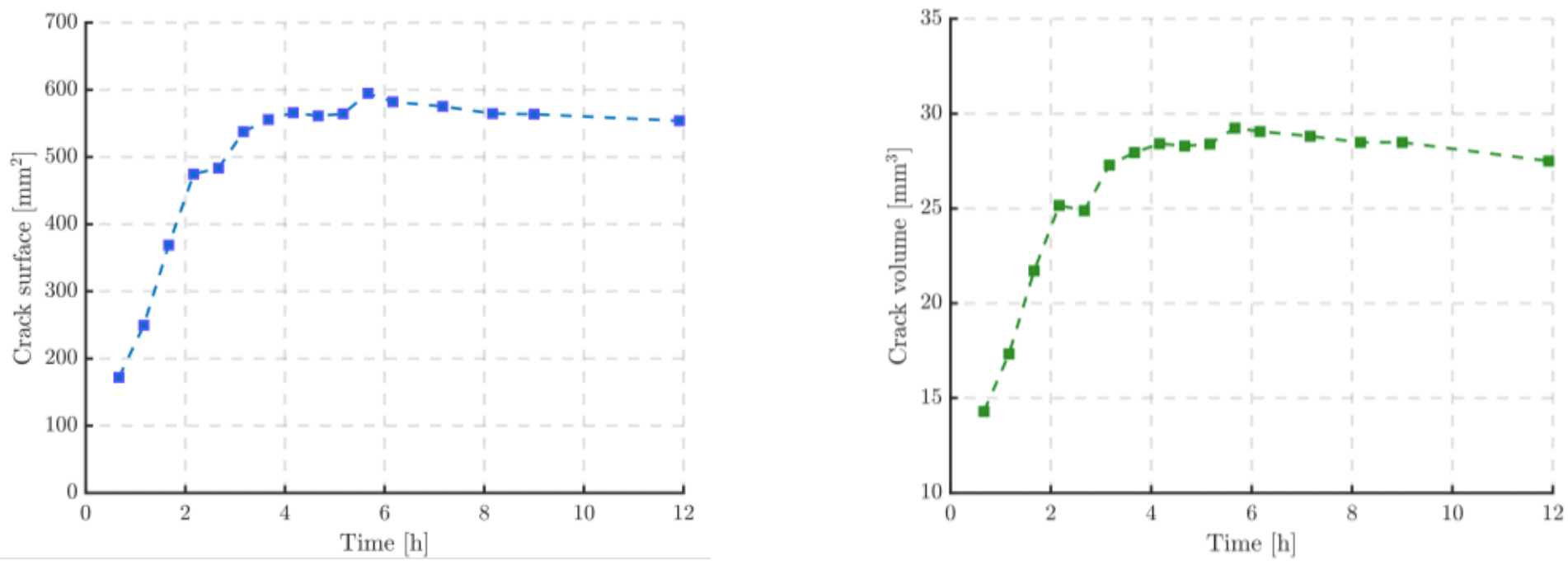

(a) (b)

**Figure 12:** (a) Crack surface development, and (b) crack volume development during the CT experiment.

Crack surface and crack volume development as presented in Figure 12 show a rapid increase in size during the first 4 hours of the experiment. We link this rapid damage progress in the sample to the onset of the second hydration phase (after the dormant period) as presented in Figure 8. Although, the temperature data shows a steady increase until the age of approx. 10 hours, the crack growth does not progress further after the age of 6 hours. Thus, this time was set as the final crack length, which does not grow any further.

This observed crack growth is more visible when presented as 3D data in the obtained ROI. In such a visualisation, one can observe pores and solid phases of the cement mortar together with the detected crack. A 3D visualisation of the detected crack is shown in Figure 13.

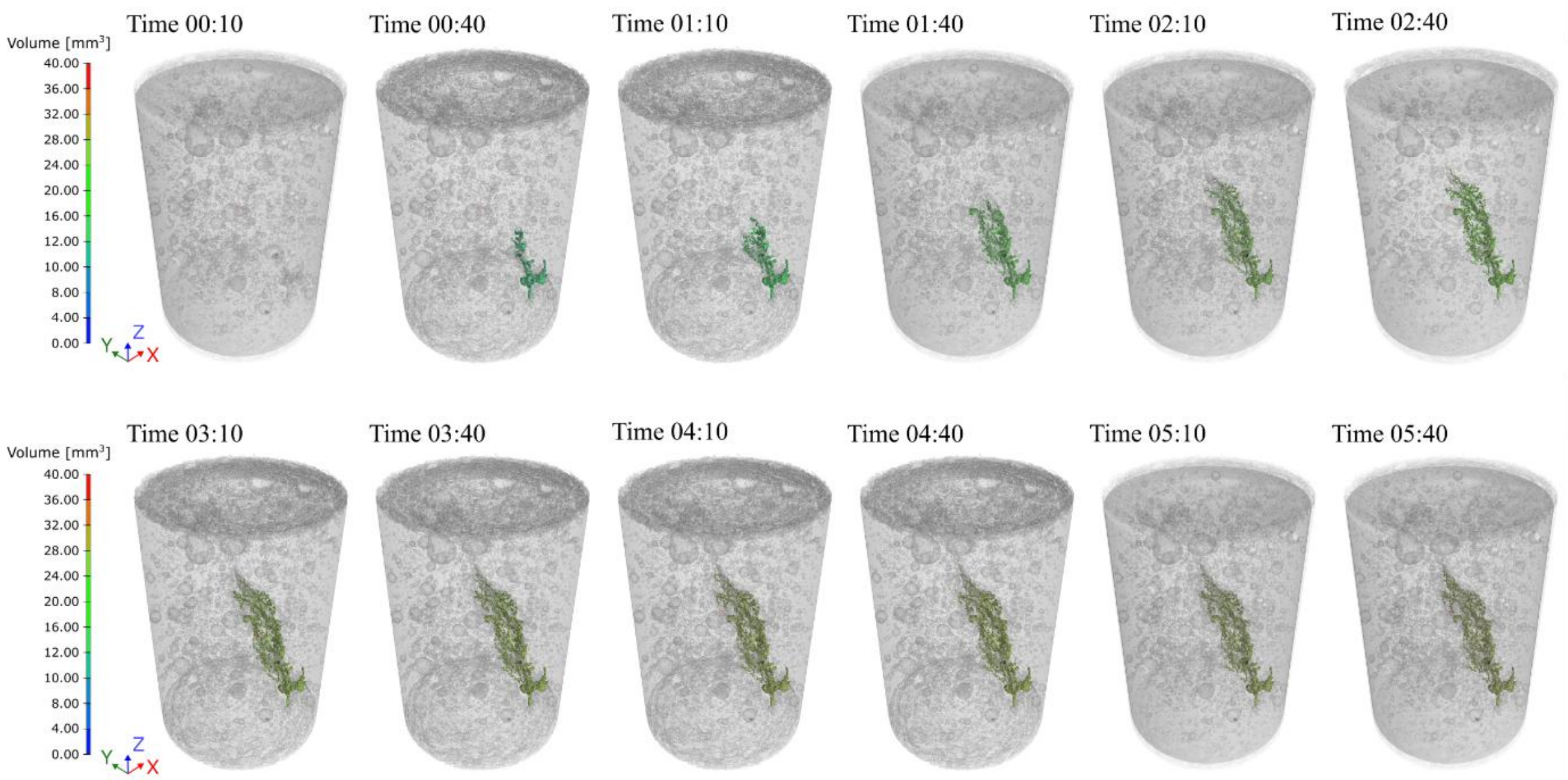

**Figure 13:** 3D visualisation of the detected crack at various ages.

Figure 13 shows an in situ crack growth throughout the experiment from 00:10 to 5:40 hours. The first CT (top left) scan shows the volume with pores without any cracks. The onset of crack growth appears at the time of 0:40 hours from the mixing of the sample, which can again be related to the second phase of cement hydration, i.e., the steady increase in the internal temperature. Another visualisation clearly presents the crack growth located inside the sample, close to the specimen centre, without any connection to the specimen's walls. The reference specimen (Sample 2 in Figure 4) placed outside of the X-ray source showed a similar crack patter and location. We do not show this tomograph for clarity's sake, as this refence sample was scanned after 25 hours without any previous μXCT scans, while the presented methodology is strictly focused on the time-lapse μXCT and early age damage localisation in the fresh cement paste.

The presence of crack within the sample is attributed to the hydration phases as well as the low w/c ratio of 0.3. This low w/c ratio leads to self-desiccation of the OPC mortar, as it is insufficient with water for cement for complete hydration (cement is consuming water form pores). This self-desiccation effect of the OPC mortar leads to a macroscopic volume change in the sample, resulting in a high strain concentration, ultimately causing crack formation [43]. Changing room temperature is another significant factor that contributes to the crack appearance in the ROI. In this case, the plastic container restrained the radial expansion of OPC mortar. Therefore, when the sample was cooled, the mortar shrank, causing damage to the MS, which was identified as a crack. To verify this statement, additional analysis is required, such as digital volume correlation (DVC), to calculate strains or to assemble simple numerical finite element models (FEM) to generate stress and strain fields. Unfortunately, the separation of these two phenomena is challenging.

An interesting observation can be made from Figure 13, in which the crack grows diagonally towards the top side of the specimen. After 5:40 hours of the sample's age, the crack growth stops, or the detected growth is insignificant. Every crack detected in the μXCT data can be located in the ROI space. The final crack size at 06:10 h age is shown in Figure 14.

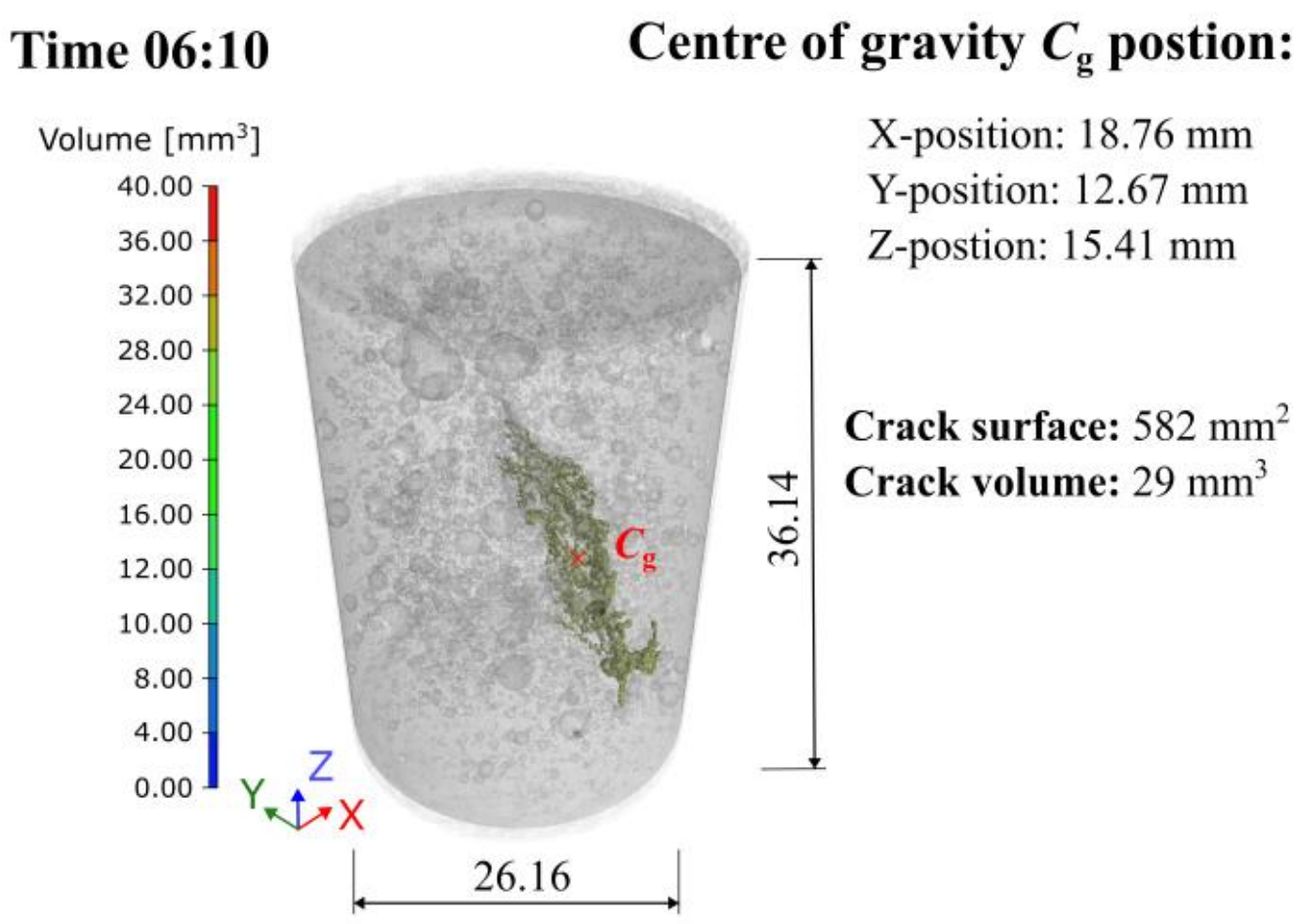

**Figure 14:** Final crack size at 06:10 hours visualized as 3D CT data with origin of used coordinate system and crack characteristics.

The final crack size at 6:00 hours, as presented in Figure 14, shows an interconnection of multiple pores with a crack growing in the diagonal direction towards the top edge of the sample. Figure 14 depicts the location of the detected crack's centre of gravity, which can be used to determine the precise position of the crack within the ROI. The measured maximum crack surface and volume were 582 mm$^2$ and 29 mm$^3$, respectively. The measured porosity, crack surface, and volume are presented in more detail in Appendix A – Experimental Details.

Additionally, the damage analysis detects the crack throughout the experiment, and it provides the information about the coordinates in Cartesian coordinate system of selected ROI. The identified coordinates of the crack's centre of gravity $C_g$ are complemented with the crack surface and crack volume.

The identification of the crack coordinates on every μXCT scan obtained in this study shows the direct location of the crack's centre of gravity $C_g$. The obtained coordinates of the centre of gravity $C_g$ throughout the experiment, can be interpreted as a crack movement in time. The overall position of the $C_g$ centre of gravity during the μXCT experiment is presented in Figure 15.

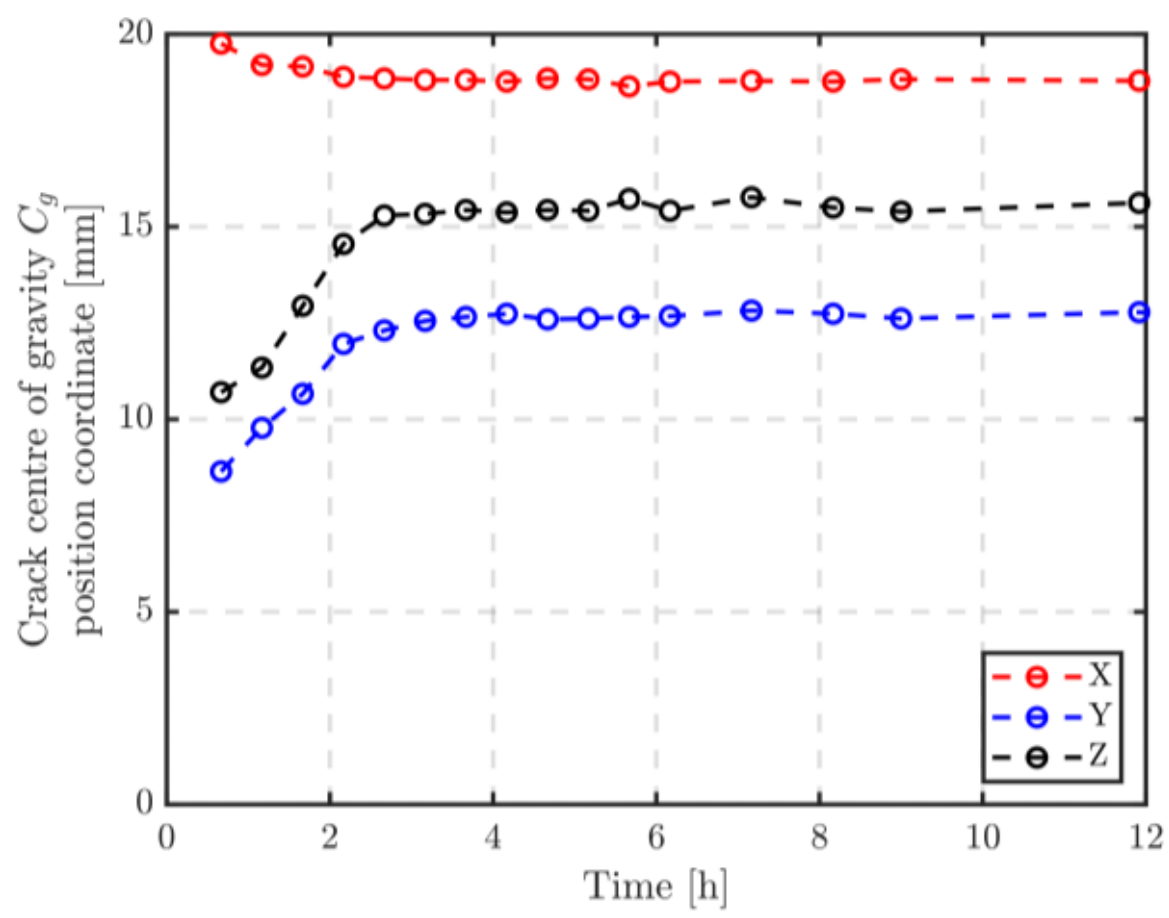

**Figure 15:** Position of centre of gravity $C_g$ of identified crack.

The results of the crack movement from Figure 15 confirm the observation made from the 3D crack visualisation (Figure 13). The crack grows in the *Y-Z* plane in the diagonal direction with a similar gradient of both coordinates. This diagonal crack growth reveals uneven temperature profiles produced by the OPC mortar hardening and mainly by the X-ray source, which could be linked to the sample rotation in the CT test setup during the CT data acquisition.

The information about the crack's centre of gravity $C_g$ could be used to characterise the crack growth from the start of the μXCT scanning, or to describe the crack growth between each μXCT scan. This information of the increase in the crack size from the onset of the fracture is shown in Figure 16(a), while Figure 16(b) shows the increment in crack size between individual μXCT scans.

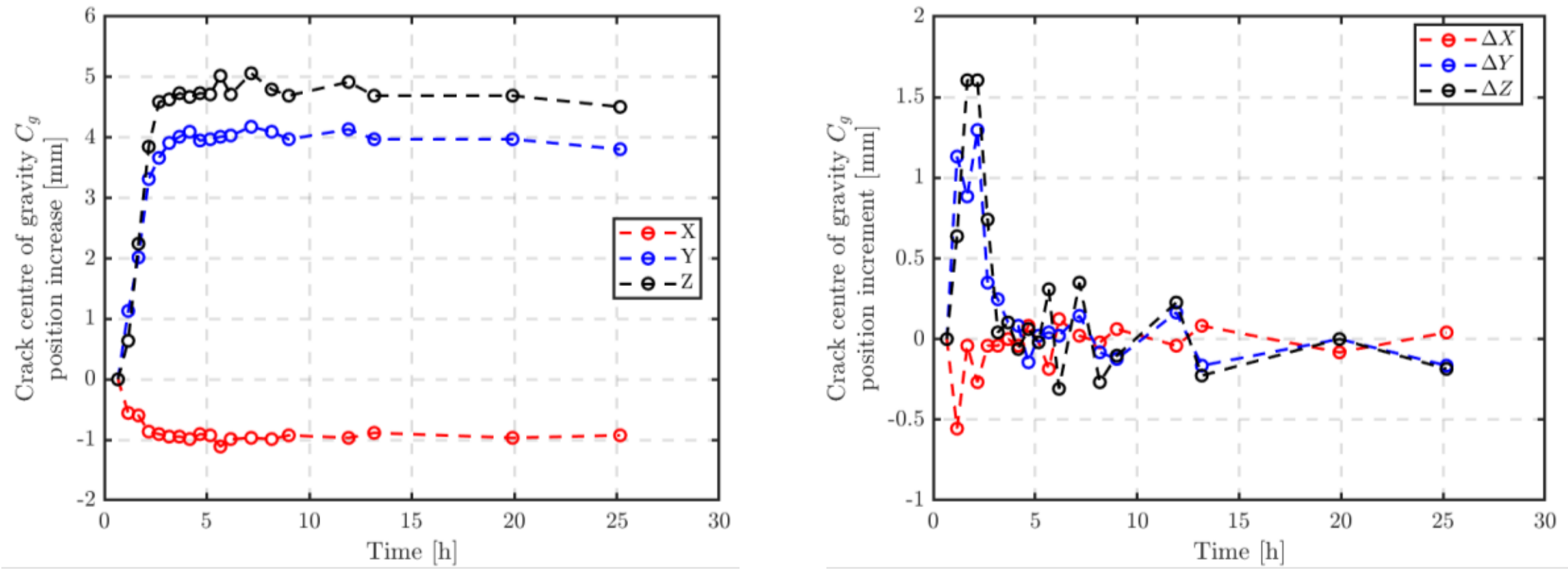

(a) (b)

**Figure 16:** (a) Crack coordinates increase from initial size, and (b) crack increment between each CT scan.

The results of the crack's centre of gravity $C_g$ coordinates in Figure 16 confirm the previous observations and give a more accurate confirmation of the diagonal crack growth, as the increment of the crack is nearly identical for both *Y* and *Z*-coordinates. The crack location increment in Figure 16(b) confirms again the observation of the crack growth made with connection to the temperature data, which shows the increase in crack size in the period from 0:40 to 2:40 hours; this is considered the second phase of cement hydration. The negative value of the X-coordinate indicates a possible interconnection of crack with pores, which produces biasing from a common planar crack shape to a spatial crack.

The connection of the observed crack with the measured temperature data is relating the crack growth to the OPC hydration phase, whereas it does not consider the shrinkage of the OPC mortar from self-desiccation as the water is consumed by OPC hydration. Nonetheless, the presented proof-of-concept methodology for material damage localisation using in situ μXCT scanning provides a valuable insight into the inner cracking process initiated by the natural process of exothermic cement hydration, which can be applied on various materials in the future.

## 3.4 Limitations

The presented results are based on the observations provided by the μXCT scanning methodology coupled with temperature acquisition during OPC mortar hardening. This study shows that capturing damage (crack) in fresh cement mortar is possible with the current experimental methods. However, there are some limitations which should be presented and discussed:

a) Environmental conditions:

Environmental conditions have the highest influence on the results as the measured temperature data (Figure 7 and Figure 8) show high temperature fluctuations during CT data acquisition. Although this additional heat was added to the analytical model, the results of μXCT analysis indicate that the temperature difference is the governing mechanism of damage initiation. Besides temperature, the humidity should be controlled too, which could reduce the chance of shrinkage of the mixture.

b) Material type

If the modern binder with supplementary cementitious materials (SCMs) were used, the SCMs would significantly influence the hydration heat released during mixture hardening. This could lead to temperature profiles diverse from the currently measured ones, resulting in a different mechanism of failure. Additionally, if, for example, alkali activated materials (AAM) were studied, a special attention would need to be given to the heat release during hardening as these mixtures have different reaction kinetics and are more prone to forming micro-cracks in the material's MS.

c) Absence of shrinkage measurement

The lack of shrinkage measurement during the experiment does not provide a clear conclusion to this experimental study. However, the acquired CT data could be used in the digital volume correlation (DVC) method, which can calculate strains in 3D. Such results would be more valuable than the ones acquired using the classic shrinkage measurement methods.

d) Time length of CT data acquisition & Spatial resolution of μXCT

The time length of the CT data acquisition directly related to the sample dimensions as well as to the parameters of the X-ray source and detector. CT data acquisition time could be reduced by reducing the dimensions of the sample, or by using a different type of X-ray source. However, reducing the dimensions would show a different thermomechanical behaviour of the sample. Both parameters are influencing the given porosity measurements. On the other hand, reducing the acquisition time by half could provide a focus on the first stage of the OPC hydration.

e) Absence of external strain measurements

An absence of external/additional strain measurements limits the study to the strain calculations from the CT data, which is highly time-consuming. The measurement of strain in fresh OPC mortar – as well as strain measurement on other samples which would increase the experimental cost – is not feasible in this type of experiment.

Besides the above-mentioned limitations, this study presented a proof-of-concept methodology allowing for the early age OPC mortar and demonstrated interesting results indicating early age damage that is progressive in time. Additionally, if this sample were used to perform an in situ compressive test with CT scanning, the results could be linked to the observed damage originating from handling and mixing the sample. Moreover, if the obtained data were used as an input to digital volume correlation (DVC), a displacement and strain fields in 3D would be calculated. Such strains would provide more insight into the damage initiation at an early age of OPC mortar. However, if such results were presented within this methodological study, the length of this paper would exceed the recommended length.

# Conclusions

This experimental study revisited the hydration of Portland cement, this time focusing on the internal changes in the material. Samples of cement mortar were made and placed in the X-Ray μCT imaging setup. μXCT images were taken during the mortar hardening, and hydration heat was measured over 25 hours of cement hydration. Prior to μXCT scanning, the adiabatic hydration heat was measured, and the chemical and mineralogical analysis of OPC was determined. With this, we were able to create a timelapse of the damage evolution in the inner structure of the material.

The measured hydration heat was compared with the analytical prediction, which was calculated based on the chemical composition of Portland cement. A relatively good agreement was obtained between the experimental and analytical predictions. This verified the standard behaviour of cement hydration and the consequent cement hardening. Additionally, an X-ray heat source was added to the development of the hydration heat, which increased the accuracy between the analytical and experimental values.

Post-processing of μXCT data allowed for separating the material phases, the sample's porosity, and, most importantly, separating the pores from the crack. Therefore, porosity was measured at various ages from the mortar mixing; the obtained value of 1.29 % corresponds to the standard value of porosity concrete. The observed scatter in the porosity measurement is due to the drying of wet cement mortar, which influenced the values of grayscale threshold used in the pore detection.

Moreover, post-processing of μXCT data localised a crack that was increasing in size from 0:40 to 6:10 hours, after which the crack size change was insignificant. The crack analysis was provided with the location of the centre of gravity $C_g$, which was changing during the μXCT experiment. Hence, the change in the centre of gravity $C_g$ of the crack could be interpreted as the crack growth. Furthermore, the crack surface and crack volume were calculated throughout the μXCT experiment. The internal crack growth was related to the hydration temperature, which allowed for a comprehensive understanding of the studied process.

The experimental results can contribute to the understanding of the damage initiation caused by the natural process of Portland cement hydration as well as to the understanding of the shrinkage behaviour of the mortar.

## Acknowledgements

Financial support provided by the Czech Science Foundation provided material under project no. 21-08772S μXCT equipment and consequent μXCT data post-processing under the project no. 23-05128S.

This paper was created as a part of project No. CZ.02.01.01/00/22_008/0004631 Materials and technologies for sustainable development within the OP JAK Program financed by the European Union and from the state budget of the Czech Republic.

## CRediT Authorship Contribution Statement

**Petr Miarka:** Visualization, Investigation, Funding acquisition, Conceptualization, Writing – original draft. **Daniel Kytýř:** Visualization, Data curation, Writing – review & editing, Funding acquisition. **Petr Koudelka:** Methodology, Visualization, Data curation, Writing – review & editing, Validation, Funding acquisition, Formal analysis. **Vlastimil Bílek:** Supervision, Formal analysis, Funding acquisition, Writing – review & editing.

## Data Availability

The data used in this study is available at: https://zenodo.org/doi/10.5281/zenodo.10390882

Supplementary material available at: https://i.imgur.com/sDWozsW.mp4

# Appendix A – Experimental Details

The measured dimensions and the total volume of the specimens are shown in Table A1 - 1.

**Table A1 - 1:** Measured dimensions and total volume.

| | Diameter *D* [mm] | Length *L* [mm] | $h_1$ [mm] | Volume [mm$^3$] |
|---|---|---|---|---|
| Reference sample | 28.9 | 144.74 | 25.6 | 78 152.55 |
| Test sample | 28.9 | 142.8 | 19.0 | 81 209.38 |

Table A1 - 2 shows a comprehensive overview of the measured porosity, the detected crack surfaces, and the volumes throughout the experiment.

**Table A1 - 2:** Pore volume, the corresponding porosity, crack surface, and crack surface obtained by μCT.

| Duration [h] | Pore volume [mm$^3$] | Porosity [%] | Crack volume [mm$^3$] | Crack surface [mm$^2$] |
|---|---|---|---|---|
| 0:30 | 264.34 | 1.37 | 14.29 | 171.86 |
| 1:00 | 227.41 | 1.18 | 17.33 | 249.53 |
| 1:30 | 227.64 | 1.18 | 21.71 | 368.79 |
| 2:00 | 229.23 | 1.19 | 25.16 | 474.53 |
| 2:30 | 258.22 | 1.34 | 24.88 | 483.49 |
| 3:00 | 269.62 | 1.40 | 27.29 | 537.64 |
| 3:30 | 233.68 | 1.21 | 27.95 | 555.75 |
| 4:00 | 235.78 | 1.22 | 28.41 | 565.93 |
| 4:30 | 272.79 | 1.41 | 28.29 | 561.10 |
| 5:00 | 271.59 | 1.41 | 28.40 | 564.32 |
| 5:30 | 237.77 | 1.23 | 29.24 | 594.80 |
| 6:00 | 238.36 | 1.23 | 29.05 | 582.10 |
| 6:30 | 237.89 | 1.23 | 28.81 | 575.31 |
| 7:00 | 273.99 | 1.42 | 28.48 | 564.62 |
| 8:00 | 236.94 | 1.23 | 28.48 | 563.73 |
| 9:00 | 265.52 | 1.37 | 27.50 | 553.84 |
| 12:00 | 263.37 | 1.36 | 27.02 | 540.77 |
| 20:00 | 273.09 | 1.41 | 28.48 | 567.13 |
| 25:00 | 233.17 | 1.21 | 25.12 | 513.62 |